\documentclass[11pt, english]{article}
 
\usepackage[a4paper,
            bindingoffset=0.2in,
            left=0.8in,
            right=0.8in,
            top=0.8in,
            bottom=0.8in,
            footskip=.25in]{geometry}

\usepackage[colorlinks=true, linkcolor=blue,citecolor=blue, allcolors=blue]{hyperref}
\usepackage[utf8]{inputenc}
\usepackage[T1]{fontenc}
\usepackage{mathtools}
\usepackage{graphicx}
\usepackage{float}
\usepackage{pdflscape}
\usepackage{color}
\usepackage[font=footnotesize,labelfont=bf]{caption}
\usepackage{amsthm}
\usepackage{amssymb}
\usepackage[colorinlistoftodos]{todonotes}
\usepackage{subfiles}

\usepackage{todonotes}
\usepackage{algorithm}
\usepackage{algpseudocode}
\usepackage[style=nature,backend=biber]{biblatex}
\usepackage{xcolor}

\newenvironment{abstracttwo}{%
  \par\nobreak\noindent
  \textbf{\textit{Abstract}\hrulefill}\par\nobreak
  \small
  \noindent\ignorespaces
}{%
  \par\nobreak\normalsize
  \vskip-\ht\strutbox\noindent
  \textbf{\hrulefill}%
}

\title{Talent is Everywhere, Mobility is Not: Mapping the Topological Anchors of Educational Pathways}
\date{}
\usepackage{authblk}

\author[1,3]{Francisco Ríos}
\author[1,3]{Fernanda Muñoz}
\author[2,3]{Valeria Bravo}
\author[3]{Gonzalo Castillo}
\author[3]{Inti Núñez}
\author[1,*]{Jorge Maluenda-Albornoz}
\author[1,4,*]{Carlos Navarrete}

\affil[1]{Departamento de Ingeniería Industrial, Facultad de Ingeniería, Universidad de Concepción}
\affil[2]{Unidad de Análisis y Calidad, Facultad de Ingeniería, Universidad de Concepción}
\affil[3]{Unidad de Innovación, Facultad de Ingeniería, Universidad de Concepción}
\affil[4]{Millennium Nucleus for the Study of Politics, Public Opinion and Media, Chile}
\affil[*]{Corresponding Authors: jorgemaluenda@udec.cl and cnavarretel@udec.cl}

\DeclareCiteCommand{\autocite}[\mkbibparens]
  {\usebibmacro{prenote}}
  {\usebibmacro{citeindex}%
   \printtext[bibhyperref]{\usebibmacro{cite}}}
  {\multicitedelim}
  {\usebibmacro{postnote}}
  
\DeclareCiteCommand{\autocitep}[]
  {\usebibmacro{prenote}}
  {\usebibmacro{citeindex}%
   \printtext[bibhyperref]{\usebibmacro{cite}}}
  {\multicitedelim}
  {\usebibmacro{postnote}}

\begin{document}

\maketitle

\begin{abstracttwo}
The relationship between socioeconomic background, academic performance, and post-secondary educational outcomes remains a significant concern for policymakers and researchers globally. While the literature often relies on self-reported or aggregate data, its ability to trace individual pathways limits these studies. Here, we analyze administrative records from over 2.7 million Chilean students (2021–2024) to map post-secondary trajectories across the entire education system. Using machine learning, we identify seven distinct student archetypes and introduce the “Educational Space,” a two-dimensional representation of students based on academic performance and family background. We show that, despite comparable academic abilities, students follow markedly different enrollment patterns, career choices, and cross-regional migration behaviors depending on their socioeconomic origins and position in the educational space. For instance, high-achieving, low-income students tend to remain in regional institutions, while their affluent peers are more geographically mobile. Our approach provides a scalable framework applicable worldwide for using administrative data to uncover structural constraints on educational mobility and inform policies aimed at reducing spatial and social inequality.

\end{abstracttwo}



Talented individuals are born in all regions, regardless of whether they come from urban or rural areas or their families' socioeconomic status (SES). However, their structural determinants, such as the lack of opportunities in their hometowns and inadequate access to quality education, have a major effect on their educational trajectories after secondary school \cite{boliver2011expansion,triventi2013stratification,sirin2005socioeconomic,diemer2020charting, drager2022role}. As only a small fraction of students have the opportunity to pursue undergraduate studies in their hometowns \cite{Desplaza60:online}, many are compelled to migrate from home seeking better educational opportunities. Related literature indicates that income is a relevant factor influencing students' decisions to migrate to higher education \cite{tuckman1970determinants}, and those with higher SES backgrounds generally perform better in academic settings \cite{rodriguez2020socio, rodriguez2024does}. While humans have historically migrated to seek new opportunities and respond to basic survival instincts \cite{faggian2018interregional, koch2023role}, and universities act as pipelines for regional human capital accumulation, often determining whether a region experiences a brain drain or brain gain phenomenon \cite{faggian2009universities}, modern migration is mostly driven by the desire to improve one's position in society. Evidence from the UK shows a high propensity for students to migrate after graduation and identifies five types of migration paths among students \cite{faggian2009universities}.

Although there is a rich discussion on various aspects of education, including educational pathways \cite{pallas2003educational}, university dropout rates \cite{maluenda2022early, casanova2018factors}, determinants of academic performance \cite{park1990determinants, sothan2019determinants, talib2012determinants}, and the influence of place of origin on academic achievement \cite{ren2021family, wai2024most}, research on university migration has yet to provide a complete picture of university graduates' migration \cite{kotavaara2018university}. However, there is a long tradition of research on the link between human capital and migration. Within this literature, it is important to note that selective migration does not necessarily imply a unidirectional flow from poorer to richer regions. Indeed, migration is selective based on both the individual characteristics of migrants and the conditions of the places they migrate to \cite{biagi2023evidence}. From a theoretical standpoint, migration is an investment where individuals weigh present costs against future returns \cite{sjaastad1962costs}. For high-performing but low-SES students, the friction of distance involves not just travel costs but also cultural and psychological barriers that increase the perceived risk of mobility, often outweighing the potential academic returns. For example, a relevant determinant of migration is the human capital of migrants \cite{haapanen2017more}. Highly skilled young people are also more likely to relocate multiple times in their lives \cite{faggian2007human}, but this tendency varies depending on the conditions of both their place of origin and destination. In this regard, evidence from Italy shows that better income conditions are a key determinant of high-skilled migration to locations with better opportunities, where individuals seek to earn more \cite{d2023stay}. Furthermore, in Wales and Scotland, people with higher human capital are more geographically mobile, yet differences between the two countries in economic, geographic, and social conditions create important variations \cite{faggian2007human}.

Given the well-established relationship between economic growth \cite{bawono2021human}, innovative ecosystems \cite{cantner2021entrepreneurial}, and social and human capital, societies need to enhance human and social capabilities to improve living conditions. Understanding the migration patterns of high-skilled individuals and their effects on human capital development is critical for policymakers and educational institutions alike. As economies increasingly rely on knowledge-based industries and compete in areas such as the global AI race \cite{korinek2021artificial}, the lack of high-quality data analysis to understand human mobility poses a challenge to policymakers, educational administrators, and decision-makers in implementing effective and strategic actions. While migration can provide access to improved educational opportunities \cite{rao2010migration}, it also presents significant challenges that are important to take into account in migration research. These challenges include adaptation to new environments, potential cultural mismatches, and the loss of household support networks, which can contribute to educational dropout and negatively impact students' academic performance and overall well-being \cite{van2024looking, tinto2012completing, odoardi2021can}. Nevertheless, most studies in the literature are limited by their reliance on cross-sectional data from self-reported surveys with restricted student samples, particularly centered in economically advantaged countries. The integration of administrative data with advanced data science and machine learning techniques can significantly deepen our understanding of the relationships between socioeconomic factors, migration patterns, and educational outcomes.

This study analyzes anonymized longitudinal records from nearly one million students who were enrolled in Chilean secondary and higher education institutions between 2021 and 2024, focusing on cross-regional migration and subsequent career choices. By using machine learning algorithms on this administrative data that covers Chile's entire post-secondary population, we clustered students based on family income, schooling quality, and standardized test performance. This analysis identified seven distinct student archetypes representing the full spectrum of the educational system. We subsequently correlated the college preferences of each archetype with patterns of internal migration and career choices, revealing a distinct divergence: high-performing students from low-SES backgrounds tend to remain in their home regions, whereas high-performing students from affluent families disproportionately relocate to universities in the capital. This persistent interaction between socioeconomic status and educational trajectories provides evidence that early-life inequalities continue to shape post-secondary mobility and professional pathways.

To the best of our knowledge, this is the first study to unpack and analyze students' educational trajectories from massive administrative records using machine learning and provides new perspectives for policymakers globally. We use Chile as a model system to understand how educational policies and structural dependencies of students shape educational migration and career choices. Likewise, our methodological approach can be replicable globally, enabling the development and formulation of data-driven education policies. The finding that student pathways vary significantly between clusters has important implications for policymakers and educational planners. Our evidence shows that, although talented students are distributed throughout the country, socioeconomic status, secondary school performance, and the quality of education received are significant factors influencing their future mobility patterns. Prestigious universities in Santiago (the capital city) attract a relevant number of regional students from higher socioeconomic backgrounds, while many talented students from low-SES backgrounds opt to attend local universities in their regions. Despite these institutions not consistently rank high in global university rankings, in light of the evidence, they constitute an essential milestone in training professionals to meet the needs of their local communities. 

\section*{Main}

Related literature on education in Chile has largely studied phenomena such as academic performance and the factors influencing educational outcomes \cite{cubillos2025evolving,maluenda2022early}. By analyzing student progress patterns, educators, academic planners, and policymakers have developed strategies to enhance the learning experience, increase retention rates, and support high-achieving students \cite{eather2022programmes}. However, these studies often rely on broad categorizations based on factors such as school of origin and family SES and largely depend on self-reported surveys and aggregated data \cite{sirin2005socioeconomic,pallas2003educational}. The lack of large-scale data studies on student performance has limited the ability of scholars and policymakers to assess the impact of policies shaping the country's evolving educational landscape over the past decades, including free university education and new standardized evaluation tests. This limitation is relevant, given that fine-grained data can offer more details about individual academic trajectories and customized learning experiences.

Machine learning techniques and data-driven studies have demonstrated their potential to address a wide range of social challenges beyond computer science, such as identifying factors of economic growth \cite{balland2022reprint}, and mapping trajectories of scientists \cite{jaramillo2025systematic}, to name a few. Yet, we oversee what remains a substantial opportunity to expand the application of data science techniques in education. At the intersection of these disciplines, studies have shown that clustering methods, such as $k$-means, can effectively monitor academic performance and guide strategic decision-making in various educational contexts \cite{oyelade2010application}, as well as optimize student admission processes \cite{santosa2021classification}.

Many theoretical studies have explored various aspects of post-secondary educational trajectories. A strand of literature focuses on college retention, revealing stark contrasts between undergraduate programs: while some maintain retention rates as high as 80\%, others see rates drop to an alarming 30\%. In Chile, university dropout rates reach 23.1\% during the first year of studies. Scholars have linked this disparity, among other factors, to students’ family backgrounds \cite{villalobos2021programas,behr2020dropping,bernardo2020}. Moreover, the emotional toll of undergraduate studies can be profound, with many students experiencing feelings of failure, frustration, and anxiety, especially in the first years of college \cite{behr2020dropping,wingert2022mindfulness}. Given these challenges, universities invest significant effort in implementing strategies to improve student retention, particularly by supporting those at risk of academic difficulties—one of the primary drivers of dropout. However, with limited resources, it is essential to focus interventions toward the students most likely to leave \cite{tinto2012completing}. Therefore, early identification of at-risk students is crucial for providing timely support and preventing dropout before it becomes irreversible. No college measure can be truly effective if it overlooks this fundamental factor, as well as understanding the sociodemographic origins of students.

Building on previous work \cite{diaz2020variables,munoz2024evaluacion,diaz2024predictor,rodriguez2019dropout}, we employ unsupervised learning clustering approach to move beyond traditional public-private educational taxonomies. We are interested in identifying student groups based on multidimensional factors--including academic performance, educational quality, and socioeconomic background--revealing patterns overlooked by conventional and rigid classifications. Furthermore, the methodology’s cross-cultural applicability highlights the transformative potential of computational social science in education research. By identifying distinct student archetypes, we enable targeted policy interventions that optimize limited resources, especially in developing nations where socioeconomic background exacerbates educational differences \cite{sirin2005socioeconomic}. This data-driven approach promotes equitable outcomes through precise student population analysis.

We identified seven distinct clusters. The names assigned to each cluster are illustrative and serve to provide a clear and comprehensible representation of the student groups within each cluster. For simplicity, we will use these archetype names as shorthand to refer to each specific cluster throughout the discussion.

\paragraph{The All-Around Achievers (Achievers)}
This cluster represents students with exceptional performance in all the metrics evaluated. Their language and math skills are exceptional, and their GPA is the highest. Furthermore, they attend high-performance schools and come from wealthy families. 

\paragraph{The Strivers (Low-Income Talent)}
This cluster groups students with high academic performance in all areas evaluated. Their language and math skills are outstanding and their GPA is high. In addition, they attend high-performance schools. However, the main difference with respect to the previous group corresponds to the SES, which is in most cases from families with relatively low incomes, showing that academic talent can flourish even in challenging socio-economic environments.

\paragraph{The Atypical Academic}
This cluster presents students with an atypical pattern: a high overall GPA, comparable to the performance of all-around achievers and low-income talent, but average performance in language and math skills. These students come from schools with relatively low performance in language and mathematics, and from families with average incomes. 

\paragraph{The Privileged Average}
This cluster groups students with average academic performance in language and math standardized tests, and a GPA slightly above average. They receive high-quality education and come from wealth families. Overall, this group represents students who benefit from a favorable educational and socioeconomic environment.

\paragraph{The Challenged Environment Academic}
This cluster is characterized by students with moderate academic performance. Although their language and math skills are in an average range, and their GPA is slightly below average, they attend high-performing schools. However, they come from families with low incomes. This group demonstrates that the quality of school education can have a positive impact, even on students from disadvantaged socioeconomic backgrounds.

\paragraph{The Resilient Underperformer}
This cluster includes students with still low academic performance. Despite their difficulties in language and mathematics, and an average GPA, these students come from families with significantly higher incomes. This suggests that despite academic disadvantages, they have a socioeconomic environment that could offer them some resilience.

\paragraph{The Disadvantaged}
This cluster groups students who consistently show the lowest performance in all evaluated metrics. Their percentiles in language skills, mathematics, and overall GPA are the lowest, indicating significant academic difficulties. Furthermore, they come from schools with relatively low performance and from families with the lowest incomes. This group faces multiple barriers that affect their educational performance.

The centroids of each cluster are shown in Figure \ref{fig:1}A, while the number of students in each cluster is presented in Figure \ref{fig:1}B.

\subsection*{The Educational Space: Mapping Student Pathways}

To operationalize the intersection of academic merit and social origin, we introduce the Educational Space: an empirical topology of opportunity. Rather than viewing career choice as a purely individual decision optimized for utility, this space reveals how location and class create \textit{bounded rationality}---where students make decisions based only on options visible within their socioeconomic constraints. We use Principal Component Analysis (PCA) to project multidimensional student characteristics into a two-dimensional plane, not merely as a visualization, but to map the structural rigidities of the Chilean educational system. This approach allows us to observe how the system could potentially reinforce barriers to intergenerational mobility by correlating a student's position in this space with their probability of spatial mobility.

We conceptualize this space as an enabler to examine how students' trajectories are projected not only by academic merit but also by the constraints imposed by their social environments. We challenge the notion of career choice as purely an individual decision, instead showing how location in this space is a determinant of academic trajectories after high school. As such, this approach serves as a cost-effective and insightful lens through which to visualize how an education system can either reinforce or mitigate barriers to intergenerational mobility.

\begin{figure}[h!]
    \centering
    \includegraphics[width=1\linewidth]{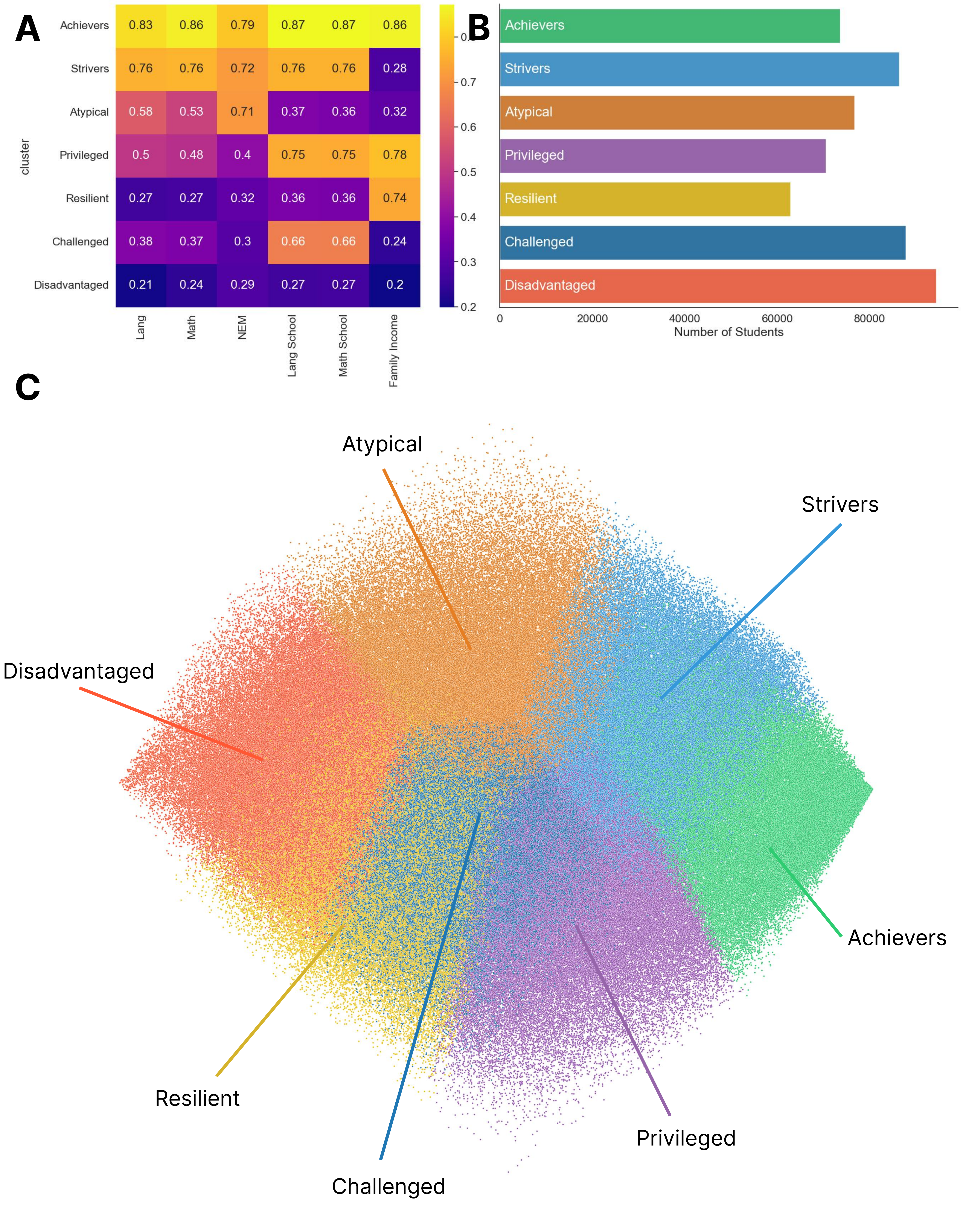}
    \caption{A. Centroid-representation of each cluster B. Total cluster composition C. The Educational Space of Chile's Post-Secondary Students.}
    \label{fig:1}
\end{figure}

Figure \ref{fig:1} reports the educational space in Chile (2021-2023), where students are colored according to their assigned clusters. Notably, distinct and well-defined regions can be traced, despite the fact that clustering was performed independently using $k$-means. This spatial distinction is not an artifact of the PCA projection but rather reflects an inherent structure in the data, reinforcing the notion that the educational space captures meaningful and systematic patterns in student characteristics.

What can we learn from the educational space? 

Consider two students with equally high scores on the university entrance exam--one from the Strivers and another from the Achievers cluster. The literature widely acknowledges that their educational trajectories are unlikely to be the same \cite{sirin2005socioeconomic}. We can note that the primary dimensions of variation in the student population correspond to academic performance (x-axis) and family socioeconomic background (y-axis), validating the space as a meaningful representation of student archetypes. Although PCA is unsupervised and does not use labeled inputs, the resulting configuration naturally reflects these well-established drivers of educational outcomes. This alignment, emerging without explicit supervision, suggests an underlying organization within the system that mirrors long-observed dynamics in the field.

\begin{figure}[h!]
    \centering
    \includegraphics[width=1\linewidth]{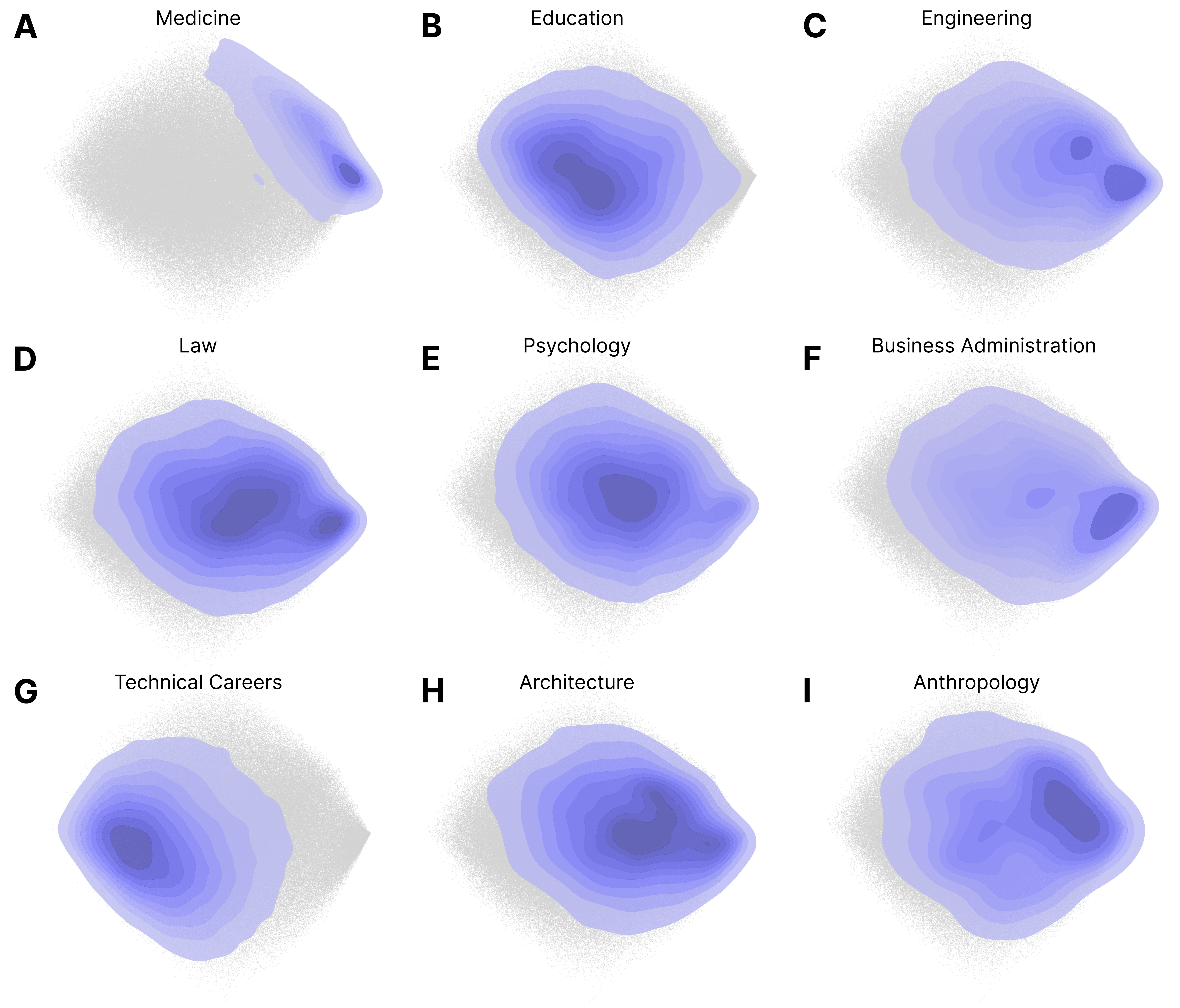}
    \caption{Density maps showing the distribution of careers in the educational space, highlighting the concentration of specific groups in particular disciplines. A) Medicine, B) Education, C) Engineering, D) Law, E) Psychology, F) Business Administration, G) Technical Careers, H) Architecture, I) Anthropology. Visualizations were created using Seaborn.}
    \label{fig:heatmap}
\end{figure}

Applying $k$-means clustering and PCA to student data reveals interesting patterns in educational trajectories. Figure \ref{fig:heatmap} presents density maps of first-year higher education enrollment for the years 2023 and 2024, segmented by career choice. By mapping students into a two-dimensional space and integrating performance metrics with enrollment data, we uncover nuanced relationships between academic achievement and career selection. For instance, medical students are predominantly found in the left periphery of the space, indicating a performance-driven pathway (Panel A). In contrast, students enrolled in technical careers tend to cluster on the right, suggesting different background and performance profiles (Panel G). Particularly illustrative are disciplines such as anthropology (Panel I) and architecture (Panel H), which tend to attract students from privileged backgrounds who also demonstrate high academic performance. This suggests that even seemingly merit-based career choices are deeply influenced by socioeconomic factors—fields perceived as more ``academic'' may be effectively inaccessible to first-generation college students due to cultural, financial, or geographic constraints. 

Equally revealing are the patterns in other disciplines. Law students show two main areas of concentration: one among Achievers and another spanning the intersection of Strivers, Privileged, and Challenged clusters (Panel D). This distribution implies that while high performance remains important, other pathways—potentially influenced by institutional access or family support—also lead students into the legal field.

These results challenge simplistic narratives of career choice based solely on merit and academic performance, aligning with existing work that considers that professional trajectories emerge from a complex interplay of academic performance, social capital, and systemic opportunities \cite{pallas2003educational,bourdieu2018forms,faggian2009universities}. The educational space not only reveals individual career paths, but also provides a methodological framework to understand the multidimensional nature of educational decision-making.

Together, these visualizations reinforce the idea that the educational space not only captures academic ability and socioeconomic background but also serves as a predictive map of career trajectories. It demonstrates that what appears to be individual agency in career decisions is often structured by broader social patterns and constraints. Its structure not only gives us the intuition to unpack well-known socioeconomic and academic divides, but also opens new avenues for exploring how students are located in the educational system. In the following sections, we leverage this representation to examine two relevant aspects in the literature, such as migration patterns and post-secondary career choices, offering deeper insights into the dynamics of educational mobility.



\subsection*{Why Clusters Are Related to Career Choices}

Several hypotheses have been proposed to explain why students choose specific career paths \cite{kniveton2004influences,duffy2007most}. A widely accepted explanation is the influence of role models, as students often shape their career decisions based on the advice of individuals they admire or aspire to emulate, high school teachers, and mentors \cite{koech2016factors,quimby2006influence,mishkin2016career}. Now, we examine these decisions through the lens of our cluster analysis, focusing on two key dimensions: enrollment patterns in higher education and cross-regional migration.

\subsubsection*{Enrollment in Higher Education}

\begin{figure}[h!]
    \centering
    \includegraphics[width=1\linewidth]{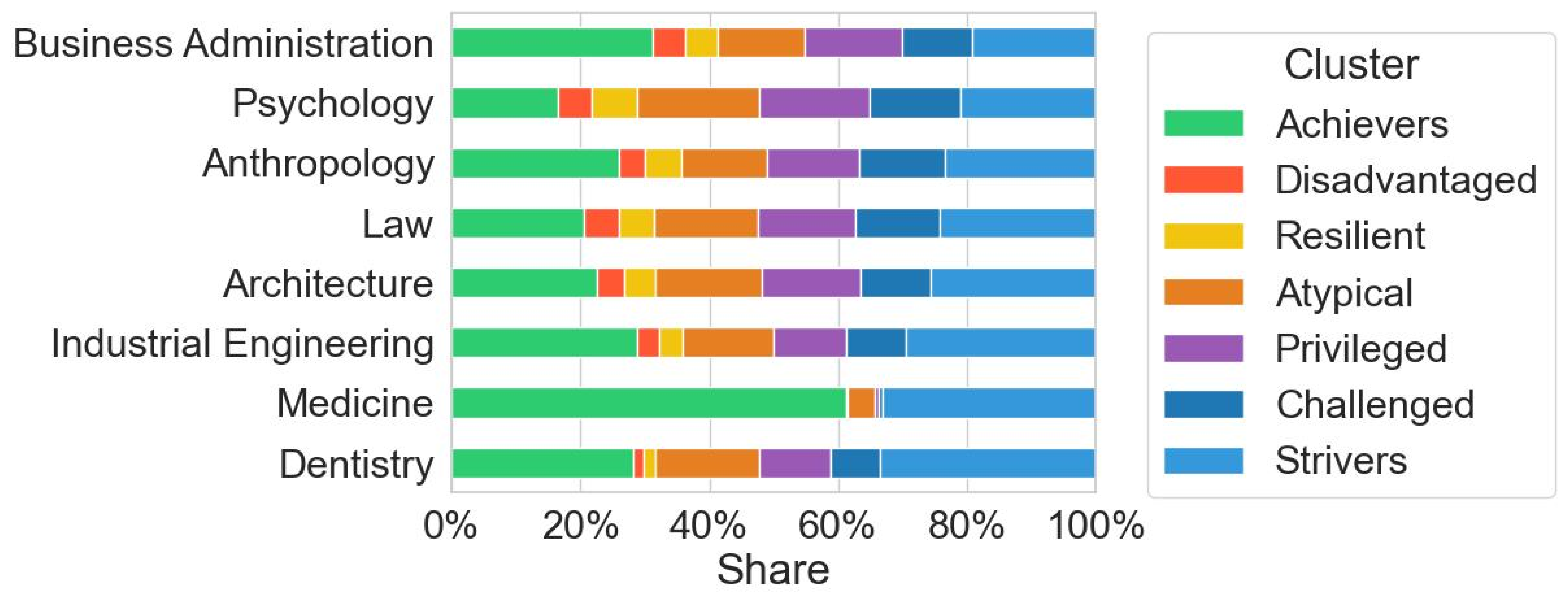}
    \caption{Distribution of student clusters across eight undergraduate programs.}
    \label{fig:3}
\end{figure}

Figure \ref{fig:3} reports the distribution of student clusters across eight undergraduate programs: Business Administration, Psychology, Anthropology, Law, Architecture, Industrial Engineering, Medicine, and Dentistry. A notable pattern emerges in the case of medicine, where nearly 60\% of students belong to the Achievers cluster, compared to only 25\% in dentistry. Nevertheless, this composition difference is marginal for Strivers. This discrepancy is particularly striking given the marginal differences in the admission cut-off scores for both programs. The contrasting cluster compositions may therefore reflect differences in the perceived value or societal prestige attributed to each career path--an aspect we hypothesize to vary according to the social group to which an individual belongs. While dentistry can be seen as valuable for some clusters, the story can be different for other ones.

We also observe substantial differences in program selection that correlate with students' socioeconomic backgrounds and their respective cluster classifications. In the Chilean context, low-income students, particularly those in the Strivers cluster, tend to favor specialized programs associated with higher projected salaries and stronger labor market demand. For example, nearly 50\% of Strivers opt for management-related fields, compared to only 30\% of Achievers. In contrast, this pattern is reversed in social science programs such as anthropology and psychology, where strivers are notably underrepresented in comparison with careers like industrial engineering. These findings suggest that career choices are shaped by differing motivations, particularly in relation to economic expectations after graduation. Nonetheless, further research is required to identify causal mechanisms underlying these patterns. However, as Hastings et al. suggest \cite{hastings2013some}, the economic return to a degree is highly heterogeneous and dependent on the specific institution in Chile.

\subsubsection*{Measuring Educational Migration}

Having established that enrollment patterns in higher education vary significantly across clusters, particularly where the most important differences are linked to the combination of family wealth and individual performance, we now shift our focus to cross-regional migration following secondary school. Specifically, we examine the mobility patterns of students within each cluster to assess whether academically talented individuals tend to seek institutions that offer world-class educational opportunities or whether they prioritize staying within their home regions due to career prospects or other contextual factors. This analysis operates at both the individual and cluster levels, aiming to understand how personal choices around mobility intersect with broader patterns and whether these behaviors can be meaningfully generalized.

This step in our analysis is particularly important because the clusters were intentionally designed to exclude geographic location as a feature, thereby avoiding potential endogeneity or bias in the analysis of migration decisions. In this vein, our theoretical expectation according to our previous finding is that SES is a driving force behind college migration.
 
\begin{figure}[h!]
    \centering
    \includegraphics[width=1\linewidth]{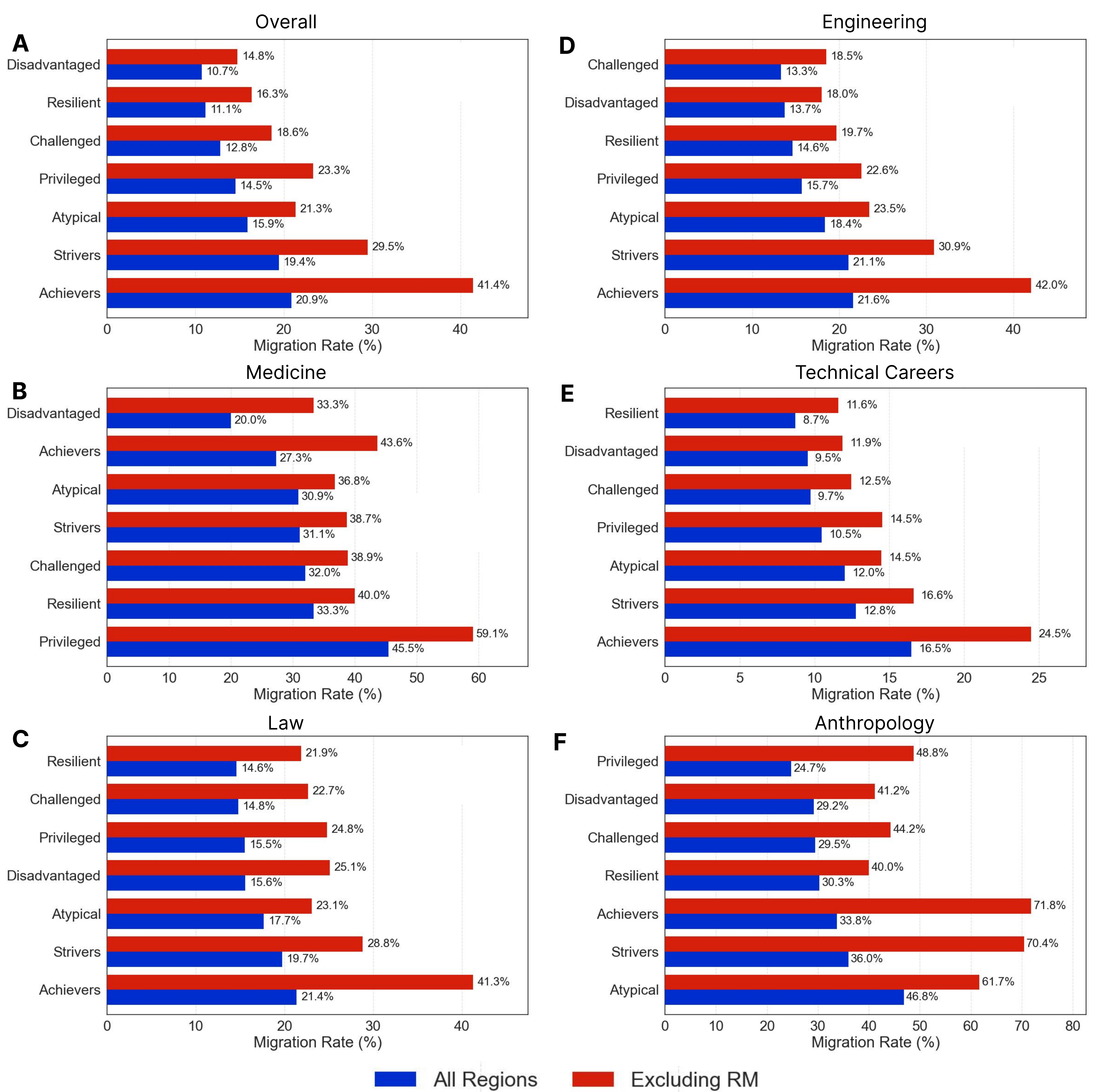}
    \caption{Cross-regional migration between 2021 and 2024. A. Full cohort. B. Medicine. C. Law. D. Engineering. E. Technical careers. F. Anthropology.
Regions on the Y-axis are ordered by overall migration rate (across all fields).}
    \label{fig:within-country-migration}
\end{figure}

Figure~\ref{fig:within-country-migration} presents cross-regional migration patterns during our study period. To analyze migration, we compared the region where students attended high school with the location of the university campus they enrolled in. Overall, we find that students in higher-performing clusters exhibit greater migration rates than those in lower-performing ones. This trend is particularly pronounced in high-value undergraduate careers. For instance, 43.6\% of students from the Achievers cluster living in regions migrated to study medicine, while only 24.5\% did so when pursuing a technical degree. 

Notably, in the case of medicine, 59.1\% of students from the Privileged cluster chose to migrate, underscoring the perceived value of this career and their willingness to relocate accordingly. These patterns shed light on a positive association between academic achievement and migration propensity, suggesting that high-performing students are significantly more likely to move in pursuit of educational opportunities. This finding resonates with theoretical frameworks on educational mobility, in which family expectations and social networks shape students' decisions to migrate \cite{faggian2007human}. Our view is that families and social networks of students take ``leaps of faith'' in supporting them to pursue studies away from home, and this trend is reinforced if the added value perceived is higher.

\begin{table}[!htbp] \centering
\begin{tabular}{@{\extracolsep{5pt}}lccccc}
\\[-1.8ex]\hline
\hline \\[-1.8ex]
& \multicolumn{5}{c}{\textit{Dependent variable: Cross-regional Migration}} \
\cr \cline{2-6}
\\[-1.8ex] & \multicolumn{1}{c}{Model 1} & \multicolumn{1}{c}{Model 2} & \multicolumn{1}{c}{Model 3} & \multicolumn{1}{c}{Model 4} & \multicolumn{1}{c}{Model 5}  \\
\\[-1.8ex] & (1) & (2) & (3) & (4) & (5) \\
\hline \\[-1.8ex]
 Language Percentile & 0.591$^{***}$ & 0.697$^{***}$ & 0.419$^{***}$ & 0.598$^{***}$ & 0.423$^{***}$ \\
& (0.029) & (0.032) & (0.056) & (0.038) & (0.057) \\
 Math Percentile & 0.460$^{***}$ & 0.513$^{***}$ & 0.226$^{***}$ & 0.414$^{***}$ & 0.231$^{***}$ \\
& (0.028) & (0.031) & (0.057) & (0.038) & (0.057) \\
 GPA Percentile & 0.437$^{***}$ & 0.471$^{***}$ & 0.455$^{***}$ & 0.463$^{***}$ & 0.454$^{***}$ \\
& (0.024) & (0.029) & (0.029) & (0.029) & (0.029) \\
 Language School Percentile & 0.364$^{***}$ & 0.529$^{***}$ & 0.523$^{***}$ & 0.524$^{***}$ & 0.521$^{***}$ \\
& (0.056) & (0.059) & (0.059) & (0.059) & (0.059) \\
 Math School Percentile & 0.379$^{***}$ & 0.547$^{***}$ & 0.539$^{***}$ & 0.538$^{***}$ & 0.536$^{***}$ \\
& (0.055) & (0.058) & (0.058) & (0.058) & (0.058) \\
 Family Income & 0.324$^{***}$ & 0.380$^{***}$ & 0.376$^{***}$ & 0.228$^{***}$ & 0.300$^{***}$ \\
& (0.020) & (0.037) & (0.037) & (0.049) & (0.052) \\
 Language × Math & & & 0.541$^{***}$ & & 0.437$^{***}$ \\
& & & (0.090) & & (0.103) \\
 Family Income × Language × Math & & & & 0.408$^{***}$ & 0.206$^{**}$ \\
& & & & (0.086) & (0.099) \\
 Atypical & & 0.106$^{***}$ & 0.145$^{***}$ & 0.163$^{***}$ & 0.167$^{***}$ \\
& & (0.037) & (0.038) & (0.039) & (0.039) \\
 Challenged & & 0.072$^{*}$ & 0.102$^{***}$ & 0.113$^{***}$ & 0.117$^{***}$ \\
& & (0.039) & (0.039) & (0.040) & (0.040) \\
 Disadvantaged & & 0.393$^{***}$ & 0.384$^{***}$ & 0.403$^{***}$ & 0.391$^{***}$ \\
& & (0.050) & (0.050) & (0.050) & (0.050) \\
 Privileged & & -0.114$^{***}$ & -0.069$^{**}$ & -0.032$^{}$ & -0.036$^{}$ \\
& & (0.029) & (0.030) & (0.033) & (0.034) \\
 Resilient & & 0.140$^{***}$ & 0.148$^{***}$ & 0.222$^{***}$ & 0.189$^{***}$ \\
& & (0.042) & (0.042) & (0.046) & (0.046) \\
 Strivers & & -0.087$^{***}$ & -0.071$^{**}$ & -0.014$^{}$ & -0.037$^{}$ \\
& & (0.029) & (0.029) & (0.033) & (0.033) \\
\hline \\[-1.8ex]
 Observations & 184659 & 184659 & 184659 & 184659 & 184659 \\
 Pseudo $R^2$ & 0.038 & 0.040 & 0.040 & 0.040 & 0.040 \\
\hline
\hline \\[-1.8ex]
\textit{Note:} & \multicolumn{5}{r}{$^{*}$p$<$0.1; $^{**}$p$<$0.05; $^{***}$p$<$0.01} \\
\end{tabular}
  \caption{Logistic Regression Models Comparison. Each dependent variable was previously standardized using Min-Max scaler. We used cluster fixed-effects.}
\label{tab}
\end{table}

Can we generalize these observations? Table \ref{tab} summarizes five logistic regression models (See Materials \& Methods), which capture the determinants of students' cross-regional migration. A consistent finding across all models is the strong positive association between individual academic performance metrics and the likelihood of migration. Specifically, higher scores in language, mathematics, and the overall GPA, as well as attending schools with higher academic standing, are all significant predictors of a student's increased propensity to migrate. Model 2, which introduces the student clusters as fixed-effect, provides a more granular view of these dynamics. Using the Achievers cluster as the baseline, we observe that belonging to certain groups is significantly associated with migration. The Atypical, Disadvantaged, and Resilient clusters exhibit a significantly higher likelihood of migration compared to the Achievers. Conversely, the Privileged and Strivers clusters do not show a statistically significant difference in migration propensity compared to the Achievers in the final, most comprehensive model.

The later models explore the interplay between academic and socioeconomic factors. Model 3 and Model 5 both include the interaction term between individual performance in math and language standardized tests, with a coefficient of 0.437 in the final model ($p<0.01$). This indicates that the positive effect of high academic amplifies the chance of an individual to migrate. Furthermore, we included a three-way interaction term (Family Income × Language × Math) in Models 4 and 5. The significant positive coefficient of 0.206 ($p<0.05$) in Model 5 implies that the combined impact of high academic ability on the likelihood of migration is not constant but is exacerbated by the student's family income. Our findings add relevant evidence to the literature that the ``returns'' to high academic achievement in terms of migration propensity are related to socioeconomic background. Model 5, thus, demonstrates that while individual academic metrics and cluster membership are powerful predictors, the decision to migrate is influenced by a complex and interacting network of a student's skills and their socioeconomic environment, which is also conditioned by their position within the educational space.

However, this pattern varies significantly across different groups. As observed, the link between academic performance and migration is substantially mediated by socioeconomic status. Students from wealthier clusters are more likely to migrate than their less privileged peers, even if their academic performance is similar. As a consequence, we argue that high-achieving students from lower-income families tend to enroll in regional universities, where we hypothesize that the presence of local support networks helps offset financial constraints. These findings highlight the dual role of regional institutions—as engines of local development and as critical educational pathways for talented students from diverse socioeconomic backgrounds \cite{sirin2005socioeconomic}.

\section*{Discussion}

Education is a cornerstone of social mobility. To design effective policies that expand opportunities for talented individuals from disadvantaged backgrounds, policymakers need reliable and data-driven evidence to guide their decisions. In this study, we analyzed millions of anonymized student records made publicly available by the Ministry of Education in Chile to unfold students' pathways, cross-regional migration, and college enrollment patterns following high school during the 2021–2024 period. To identify representative groups of students, we employed $k$-means clustering using variables such as individual performance, family wealth, and the quality of education received. This approach offers a clear advantage over traditional categorizations--typically based on school type (municipal, subsidized private, or fully private)--because it enables a more fine-grained classification grounded in performance-related characteristics rather than relying solely on socioeconomic background or institutional affiliation. To better visualize these relationships, we introduce the concept of \textit{Educational Space}, a topology that represents students’ positions in a two-dimensional landscape. Our view is that this space provides a powerful tool for analyzing future educational trajectories, as it captures the underlying interaction between individual performance and socioeconomic background.

Building on the seven student clusters, we examined first-year college enrollment and cross-regional migration patterns at both the individual and cluster levels. Our results show that higher-performing students are more likely to migrate than their lower-performing peers, a pattern that becomes even more pronounced when focusing on non-metropolitan regions. Notably, regional students from the low-SES, high-achieving cluster tend to enroll in nearby institutions at substantially higher rates than their high-SES counterparts, who are more likely to relocate to the capital city. This indicates that socioeconomic status shapes mobility opportunities even among students with comparable academic performance and emphasizes the important function of regional universities as training hubs for high-achieving students who can lack the financial resources or support networks needed to access more prestigious institutions farther from home. Despite this, enrollment patterns in fields such as medicine, anthropology, and engineering indicate that academic performance alone does not fully explain individual choices. A student's position within the educational space seems to be a significant determinant of their educational trajectory, pointing to an important avenue for further research.

Based on Bourdieu and Passeron's theory of social reproduction \cite{bourdieu2018forms}, it can be argued that educational systems tend to legitimize and perpetuate pre-existing social hierarchies by transforming differences in economic and cultural capital into seemingly natural differences in academic performance. In accordance with our findings, it is reasonable to assert that students from affluent backgrounds not only have superior material resources but also possess cultural dispositions and habits that are more congruent with the educational system's expectations, thereby easing their transition to more esteemed institutions. This structural advantage is further reinforced by symbolic mechanisms that present academic success as the exclusive outcome of individual effort, thereby obscuring the social constraints that shape educational trajectories from an early age. At this point, it is also relevant to incorporate the concept of anchored mobility \cite{sheller2006new}, which helps contextualize how the possibility of movement is mediated by economic capital, social networks, and emotional ties. Consequently, the patterns of migration and career choice observed in this study can be interpreted not only as reflections of individual abilities or aspirations but also as expressions of a system that reproduces social inequalities through differentiated educational pathways \cite{reay2018miseducation,bourdieu2018forms}.

For students from low socioeconomic backgrounds, even those with high academic achievement --particularly those in the Strivers cluster--  migration can entail a financial risk as well as a disruption of essential support networks, reducing their willingness to move to more prestigious institutions or to cities with theoretically greater opportunities. As a result, the geography of educational decisions is shaped by unequal opportunities, reinforcing territorial disparities in access to cultural and professional capital \cite{coulter2016re}. Acknowledging this dimension highlights that educational migration functions simultaneously as a resource and a constraint, and that immobility is often driven by financial frictions rather than by revealed preferences.

Likewise, differences observed in career choice among students with different socioeconomic and academic profiles can be analyzed from a more complex perspective that transcends the rational cost-benefit approach. Studies on educational aspirations suggest that vocational decisions are formed within socially constructed and unequally distributed frameworks of possibility \cite{zipin2015educating}. In contexts of vulnerability, students may develop aspirations ``adjusted'' to what they perceive as realistically achievable based on their environment, their family role models, and the symbolic representation of certain professions (bounded aspirations), limiting their options before effectively exploring their viability, which generates structural self-limitation \cite{hart2016aspirations}. This aligns with our observation from the Educational Space, which, in light of the evidence, is able to capture these adjusted aspirations. Thus, despite being talented, many low-SES students may prioritize their career choices according to this phenomenon, on the one hand, limiting themselves in terms of their chosen university, but also in terms of the degree program they select to ensure the best use of their academic performance within their limited aspirations. 

This study has important limitations that should be addressed in future research. Due to data constraints, we focused solely on the educational pathways of high school and college students enrolled in Chilean institutions during a four-year period. This may impact some of our findings, particularly those related to individuals from the highest SES backgrounds who migrate to pursue undergraduate studies abroad. Furthermore, as mentioned earlier, the methodological changes in the standardized tests of 2022 and 2023 could influence specific results. Even though our robustness checks indicate that the main trends remain consistent, introducing a new exam designed to assess more profound mathematical knowledge has decreased the number of students from low-income families being accepted into higher-ranked universities. This is a significant issue that warrants further investigation in future studies.

We hope that this paper opens new avenues for research and inspires further exploration of how machine learning can deepen our understanding of individual-level student records and enrich the education literature. We believe that the educational space is an opportunity to uncover the underlying structure of our educational systems and assess how policies align with challenges such as intergenerational mobility. This methodology offers a blueprint for other countries, especially those with similar challenges in educational inequality and data availability, to conduct evidence-based analysis and formulate policies that foster equitable access to higher education. 

One potential direction for future research is to examine students' migration patterns after completing their undergraduate studies and analyze retention rates using our clustering approach. As the literature suggests, individuals tend to migrate to locations where they already have established networks, such as family, friends, or professional connections \cite{gurak1992migration}. Building on this premise, expanding access to high-quality education for individuals from small towns could have significant long-term benefits for their hometowns, as these places are strong candidates for relocation after graduation--provided they offer adequate professional opportunities. These benefits could manifest in various ways, including the creation of new businesses, improved access to amenities, and even better healthcare services.

\section*{Materials and Methods}

\subsection*{Sample construction}

We used anonymized records of students in Chile provided by the Ministry of Education from 2021 to 2024. This data is publicly available for research purposes. The data is at the individual level, in which each individual is assigned a unique encrypted DNI to avoid reidentification. This unique identifier allowed us to combine data at different stages of their educational pathways. The datasets, along with their corresponding dictionaries, can be found in \href{https://datosabiertos.mineduc.cl/}{https://datosabiertos.mineduc.cl/}.

Due to data constraints derived from the primary source, we acknowledge that the start of our observation window (2021) coincides with the COVID-19 pandemic, which likely suppressed overall mobility. However, our analysis of the 2023 and 2024 cohorts shows that the gap in migration rates between high-SES and low-SES clusters remains significant, even as mobility restrictions are lifted. This provides evidence that the patterns observed are driven by socioeconomic anchors rather than temporary health restrictions.

The dimensions identified as relevant for this study are the following:

\paragraph{Individual Performance Assessment}
Our evaluation of individual student performance incorporates two metrics: (1) normalized high school grades based on intra-school rankings, (2) standardized reading comprehension scores, and (3) standardized mathematics scores. The reading and mathematics assessments are derived from the Prueba de Acceso a la Educación Superior (PAES), Chile's national standardized test for university admissions, comparable to the SAT in the United States. This examination has evolved over time, previously known as the Prueba de Transición Universitaria (PTU) in 2022 and the Prueba de Selección Universitaria (PSU) from 2005 to 2021. Although PAES introduced methodological changes and updated content assessment approaches in 2023, we determine that it remains the most robust and comprehensive instrument available in Chile to evaluate individual academic performance.

\paragraph{Quality of Education Received} To assess the quality of education, we employ a multistep process. First, we use each student's mean score in math and reading comprehension tests, providing a balanced measure of academic performance in standardized tests. Next, we aggregated these scores at the secondary school level, using the median student score to represent each school's overall performance. This approach mitigates the influence of outliers. Finally, we establish a school ranking metric. In cases where schools have identical median scores, we assign them the same rank, ensuring fairness in the evaluation process. This methodology offers a comprehensive, yet nuanced view of educational quality in secondary schools. This is represented as follows:

\begin{equation}
S_i = \frac{M_i + R_i}{2}
\end{equation}
where:
\begin{itemize}
    \item $S_i$ = Overall score for student $i$
    \item $M_i$ = Math test score for student $i$
    \item $R_i$ = Reading comprehension score for student $i$
\end{itemize}

\begin{equation}
SP_j = \text{median}(S_1, S_2, \ldots, S_n)
\end{equation}
where:
\begin{itemize}
    \item $SP_j$ = School performance score for school $j$
    \item $S_1, S_2, \ldots, S_n$ = Scores of all students in school $j$
\end{itemize}

\paragraph{Cross-Regional Migration} To quantify within-country student migration for undergraduate studies, we constructed a binary variable indicating whether a student's enrolled university is in the same region as their home region. This approach leverages the geographical information in our secondary school dataset, including each student's residential region. By comparing this with the location of their chosen university, we can effectively track patterns of educational migration within the country. This measure allows us to distinguish between local students pursuing higher education and those relocating for their studies. It provides valuable insights into the spatial dynamics of educational choices and potential regional disparities in higher education access and quality.

All variables were subjected to a min-max normalization procedure to facilitate comparability across variables and mitigate the impact of differing scales. This transformation linearly scaled each variable to a standardized range [0, 1], where 0 corresponds to the minimum observed value and 1 to the maximum. We perform this transformation for each year, to avoid distortions derived from the grade scale update in PAES, PSU, and PTU.

\subsection*{Clustering}
We opted for a simple clustering metric as $k$-means. This method partitions a dataset into \( k \) clusters by iteratively assigning data points to the nearest cluster centroid and updating the centroids according to the mean of the assigned points. The algorithm seeks to minimize the variance within the cluster or the sum of squared distances between the points and their centroids.

\[
J = \sum_{i=1}^k \sum_{x \in C_i} \| x - \mu_i \|^2
\]

where \( J \) is the objective function to be minimized, \( C_i \) represents the set of data points in the cluster \( i \), \( \mu_i \) is the centroid of the cluster \( i \), and \( \| x - \mu_i \|^2 \) is the squared Euclidean distance between a data point \( x \) and the centroid \( \mu_i \).

We employed two complementary techniques to determine the optimal number of clusters: the elbow method and silhouette analysis. By combining these approaches, we aim to balance the trade-off between model complexity and cluster quality, thereby ensuring a robust and meaningful clustering solution. After multiple iterations, we set up seven clusters according to the values obtained for both techniques.

\subsection*{Constructing the Educational Space}

To represent students in a shared educational space, we applied Principal Component Analysis (PCA) to a standardized set of academic and socioeconomic features: percentiles in language, mathematics, and GPA scores; school-level language and mathematics percentiles; and family income. Standardization ensures that all variables contribute equally to the analysis regardless of scale.

We retained the first two principal components that capture the dominant axes of variation among students. The resulting low-dimensional representation enables visual inspection of patterns in the data, most notably the emergence of clusters with distinct educational profiles. These two components form the basis for our clustering and subsequent analyses.

\subsection*{Estimation Model}

To analyze the factors influencing cross-regional student migration, we employ a logistic regression model, which is well-suited for estimating the probability of a binary outcome. Our dependent variable, $Y_{it}$, is a binary indicator of whether the student $i$ migrates between regions at time $t$. In order to avoid distortions from using the Metropolitan Region, we excluded those records from the model. The model, which incorporates fixed effects for student clusters belongingness, is specified as follows:

\begin{equation}
\label{eq:logit_model}
\begin{split}
\text{Logit}(P(Y_{it}=1)) &= \alpha + \beta_1 X_{1,it} + \dots + \beta_k X_{k,it} + \sum_c^{C-1} \gamma_c \cdot I(\text{Cluster}_i=c) \\
&\quad + \delta_1(X_{\text{lang}} \cdot X_{\text{math}})_{it} + \delta_2(X_{\text{faminc}} \cdot X_{\text{lang}} \cdot X_{\text{math}})_{it} + \epsilon_{it}
\end{split}
\end{equation}

In this equation, $X_{1,it}, \dots, X_{k,it}$ represents a vector of economic and academic control variables for student $i$ at time $t$. These variables include the scaled student's percentile scores in language, math and GPA scores, as well as the average percentile ranks of their school and their family income. The term $I(\text{Cluster}_i=c)$ is an indicator function that captures the fixed effects of the student's assigned cluster, with Archivers cluster serves as the reference group. The model also includes interaction terms, such as $(X_{\text{lang}} \cdot X_{\text{math}})_{it}$, to test if the effect of a student's performance in one subject is conditional on their performance in another. A three-way interaction, $(X_{\text{faminc}} \cdot X_{\text{lang}} \cdot X_{\text{math}})_{it}$, is also incorporated to examine how the combined effect of academic skills on migration is moderated by family income. The coefficients of interest, represented by $\beta$, $\gamma$, and $\delta$, are estimated using the maximum likelihood estimate. We present a series of models that progressively add these variables and interactions to demonstrate their incremental impact on explaining student migration.

\section*{Funding Statement}
Not applicable.

\printbibliography

@article{balland2022reprint,
  title={Reprint of the new paradigm of economic complexity},
  author={Balland, Pierre-Alexandre and Broekel, Tom and Diodato, Dario and Giuliani, Elisa and Hausmann, Ricardo and o'Clery, Neave and Rigby, David},
  journal={Research Policy},
  volume={51},
  number={8},
  pages={104568},
  year={2022},
  publisher={Elsevier}
}

@article{rodriguez2019dropout,
  title={Dropout and transfer paths: What are the risky profiles when analyzing university persistence with machine learning techniques?},
  author={Rodr{\'\i}guez-Mu{\~n}iz, Luis J and Bernardo, Ana B and Esteban, Mar{\'\i}a and D{\'\i}az, Irene},
  journal={Plos one},
  volume={14},
  number={6},
  pages={e0218796},
  year={2019},
  publisher={Public Library of Science San Francisco, CA USA}
}

@article{diaz2024predictor,
  title={Predictor variables of immigrant rooting: A model using machine learning},
  author={D{\'\i}az-Ram{\'\i}rez, Jorge and Berr{\'\i}os-Riquelme, Jos{\'e} and Maluenda-Albornoz, Jorge and Grau-Rengifo, Olaya and Castillo-Rozas, Gustavo and Vidal-Figueroa, Carla},
  journal={Interciencia},
  volume={49},
  number={12},
  pages={701--710},
  year={2024},
  publisher={Interciencia}
}

@phdthesis{munoz2024evaluacion,
  title={EVALUACI{\'O}N DE MODELO PREDICTIVO DEL ABANDONO UNIVERSITARIO BASADO EN CLUSTERES DE COMPORTAMIENTO DE LOS/AS ESTUDIANTES DE PRIMER A{\~N}O DE INGENIER{\'I}A UDEC},
  author={Mu{\~n}oz, Fernanda},
  year={2024},
  school={Universidad de Concepci{\'o}n}
}

@inproceedings{diaz2020variables,
  title={Variables influencing university dropout: A machine learning-based study},
  author={D{\'\i}az, Irene and Bernardo, Ana B and Esteban, Mar{\'\i}a and Rodr{\'\i}guez-Mu{\~n}iz, Luis J},
  booktitle={International Conference on EUropean Transnational Education},
  pages={94--103},
  year={2020},
  organization={Springer}
}

@article{triventi2013stratification,
  title={Stratification in higher education and its relationship with social inequality: A comparative study of 11 European countries},
  author={Triventi, Moris},
  journal={European sociological review},
  volume={29},
  number={3},
  pages={489--502},
  year={2013},
  publisher={Oxford University Press}
}

@article{boliver2011expansion,
  title={Expansion, differentiation, and the persistence of social class inequalities in British higher education},
  author={Boliver, Vikki},
  journal={Higher education},
  volume={61},
  number={3},
  pages={229--242},
  year={2011},
  publisher={Springer}
}

@article{cubillos2025evolving,
  title={The Evolving Relationship Between Reading Motivation and Achievement: A Longitudinal Study},
  author={Cubillos, Montserrat and Troncoso, Rodrigo},
  journal={Education Sciences},
  volume={15},
  number={10},
  pages={1274},
  year={2025},
  publisher={MDPI}
}

@techreport{hastings2013some,
  title={Are some degrees worth more than others? Evidence from college admission cutoffs in Chile},
  author={Hastings, Justine S and Neilson, Christopher A and Zimmerman, Seth D},
  year={2013},
  institution={National Bureau of Economic Research}
}

@article{sjaastad1962costs,
  title={The costs and returns of human migration},
  author={Sjaastad, Larry A},
  journal={Journal of political Economy},
  volume={70},
  number={5, Part 2},
  pages={80--93},
  year={1962},
  publisher={The University of Chicago Press}
}

@book{tinto2012completing,
  title={Completing college: Rethinking institutional action},
  author={Tinto, Vincent},
  year={2012},
  publisher={University of Chicago Press}
}

@article{sirin2005socioeconomic,
  title={Socioeconomic status and academic achievement: A meta-analytic review of research},
  author={Sirin, Selcuk R},
  journal={Review of educational research},
  volume={75},
  number={3},
  pages={417--453},
  year={2005},
  publisher={Sage Publications Sage CA: Thousand Oaks, CA}
}

@incollection{pallas2003educational,
  title={Educational transitions, trajectories, and pathways},
  author={Pallas, Aaron M},
  booktitle={Handbook of the life course},
  pages={165--184},
  year={2003},
  publisher={Springer}
}

@techreport{korinek2021artificial,
  title={Artificial intelligence, globalization, and strategies for economic development},
  author={Korinek, Anton and Stiglitz, Joseph E},
  year={2021},
  institution={National Bureau of Economic Research}
}

@article{hart2016aspirations,
  title={How do aspirations matter?},
  author={Hart, Caroline Sarojini},
  journal={Journal of human development and capabilities},
  volume={17},
  number={3},
  pages={324--341},
  year={2016},
  publisher={Taylor \& Francis}
}

@article{zipin2015educating,
  title={Educating for futures in marginalized regions: A sociological framework for rethinking and researching aspirations},
  author={Zipin, Lew and Sellar, Sam and Brennan, Marie and Gale, Trevor},
  journal={Educational philosophy and theory},
  volume={47},
  number={3},
  pages={227--246},
  year={2015},
  publisher={Taylor \& Francis}
}

@article{coulter2016re,
  title={Re-thinking residential mobility: Linking lives through time and space},
  author={Coulter, Rory and Ham, Maarten van and Findlay, Allan M},
  journal={Progress in human geography},
  volume={40},
  number={3},
  pages={352--374},
  year={2016},
  publisher={SAGE Publications Sage UK: London, England}
}

@article{reay2018miseducation,
  title={Miseducation: Inequality, education and the working classes},
  author={Reay, Diane},
  journal={International Studies in Sociology of Education},
  volume={27},
  number={4},
  pages={453--456},
  year={2018},
  publisher={Taylor \& Francis}
}

@misc{sheller2006new,
  title={The New Mobilities Paradigm. Environment and Planning A”: Economy and Space, 38 (2), 207--226},
  author={Sheller, M and Urry, J},
  year={2006}
}

@incollection{bourdieu2018forms,
  title={The forms of capital},
  author={Bourdieu, Pierre},
  booktitle={The sociology of economic life},
  pages={78--92},
  year={1986},
  publisher={Routledge}
}

@article{jaramillo2025systematic,
  title={Systematic comparison of gender inequality in scientific rankings across disciplines},
  author={Jaramillo, Ana Maria and Macedo, Mariana and Oliveira, Marcos and Karimi, Fariba and Menezes, Ronaldo},
  journal={arXiv preprint arXiv:2501.13061},
  year={2025}
}

@article{wingert2022mindfulness,
  title={Mindfulness-based strengths practice improves well-being and retention in undergraduates: A preliminary randomized controlled trial},
  author={Wingert, Jason R and Jones, Jeffrey C and Swoap, Robert A and Wingert, Heather M},
  journal={Journal of American College Health},
  volume={70},
  number={3},
  pages={783--790},
  year={2022},
  publisher={Taylor \& Francis}
}

@book{bernardo2020,
    author    = {Bernardo, A. B. and Herrero, E. T. and Almeida, L. S. and Pérez, J. C. N.},
    title     = {Motivos y factores explicativos del abandono de los estudios: Claves y estrategias para superarlo},
    publisher = {Ediciones Pirámide},
    year      = {2020}
}

@article{behr2020dropping,
  title={Dropping out of university: a literature review},
  author={Behr, Andreas and Giese, Marco and Teguim Kamdjou, Herve D and Theune, Katja},
  journal={Review of Education},
  volume={8},
  number={2},
  pages={614--652},
  year={2020},
  publisher={Wiley Online Library}
}

@article{villalobos2021programas,
  title={Programas proped{\'e}uticos de nivelaci{\'o}n acad{\'e}mica en favor del acceso a estudios universitarios en Chile},
  author={Villalobos Ega{\~n}a, Lenin Roberto},
  journal={Conrado},
  volume={17},
  number={82},
  pages={30--42},
  year={2021},
  publisher={Universidad de Cienfuegos.}
}

@article{eather2022programmes,
  title={Programmes targeting student retention/success and satisfaction/experience in higher education: A systematic review},
  author={Eather, Narelle and Mavilidi, Myrto F and Sharp, Heather and Parkes, Robert},
  journal={Journal of Higher Education Policy and Management},
  volume={44},
  number={3},
  pages={223--239},
  year={2022},
  publisher={Taylor \& Francis}
}

@article{bawono2021human,
  title={Human capital, technology, and economic growth: A case study of Indonesia},
  author={Bawono, Suryaning},
  journal={Journal of Asian Finance, Economics and Business},
  year={2021}
}

@article{cantner2021entrepreneurial,
  title={Entrepreneurial ecosystems: a dynamic lifecycle model},
  author={Cantner, Uwe and Cunningham, James A and Lehmann, Erik E and Menter, Matthias},
  journal={Small Business Economics},
  volume={57},
  pages={407--423},
  year={2021},
  publisher={Springer}
}

@article{haapanen2017more,
  title={More educated, more mobile? Evidence from post-secondary education reform},
  author={Haapanen, Mika and B{\"o}ckerman, Petri},
  journal={Spatial Economic Analysis},
  volume={12},
  number={1},
  pages={8--26},
  year={2017},
  publisher={Taylor \& Francis}
}

@article{van2024looking,
  title={Looking beyond primary barriers: Support workers’ perspectives on school dropout among students with a migration background},
  author={Van Den Berghe, Lana and Pouille, Aline and Vandevelde, Stijn and De Pauw, Sarah SW},
  journal={Journal of Ethnic \& Cultural Diversity in Social Work},
  volume={33},
  number={1},
  pages={34--46},
  year={2024},
  publisher={Taylor \& Francis}
}

@article{d2023stay,
  title={Stay or emigrate? How social capital influences selective migration in Italy},
  author={D’Ingiullo, Dario and Odoardi, Iacopo and Quaglione, Davide},
  journal={Regional Studies, Regional Science},
  volume={10},
  number={1},
  pages={529--548},
  year={2023},
  publisher={Taylor \& Francis}
}

@article{faggian2007human,
  title={Human capital, higher education and graduate migration: an analysis of Scottish and Welsh students},
  author={Faggian, Alessandra and McCann, Philip and Sheppard, Stephen},
  journal={Urban Studies},
  volume={44},
  number={13},
  pages={2511--2528},
  year={2007},
  publisher={Sage Publications Sage UK: London, England}
}

@article{biagi2023evidence,
  title={Evidence of self-selection and spatial mismatch in interregional migration: the case of Italy},
  author={Biagi, Bianca and Detotto, Claudio and Faggian, Alessandra},
  journal={Oxford Economic Papers},
  volume={75},
  number={3},
  pages={858--872},
  year={2023},
  publisher={Oxford University Press}
}

@article{duffy2007most,
  title={What is most important to students' long-term career choices: Analyzing 10-year trends and group differences},
  author={Duffy, Ryan D and Sedlacek, William E},
  journal={Journal of Career Development},
  volume={34},
  number={2},
  pages={149--163},
  year={2007},
  publisher={Sage Publications Sage CA: Los Angeles, CA}
}

@article{kniveton2004influences,
  title={The influences and motivations on which students base their choice of career},
  author={Kniveton, Bromley H},
  journal={Research in education},
  volume={72},
  number={1},
  pages={47--59},
  year={2004},
  publisher={SAGE Publications Sage UK: London, England}
}

@misc{rao2010migration,
  title={Migration, education and socio-economic mobility},
  author={Rao, Nitya},
  year={2010},
  publisher={Taylor \& Francis}
}

@article{kotavaara2018university,
  title={University graduate migration in Finland},
  author={Kotavaara, Niina and Kotavaara, Ossi and Rusanen, Jarmo and Muilu, Toivo},
  journal={Geoforum},
  volume={96},
  pages={97--107},
  year={2018},
  publisher={Elsevier}
}

@article{sothan2019determinants,
  title={The determinants of academic performance: evidence from a Cambodian University},
  author={Sothan, Seng},
  journal={Studies in Higher Education},
  volume={44},
  number={11},
  pages={2096--2111},
  year={2019},
  publisher={Taylor \& Francis}
}

@article{park1990determinants,
  title={Determinants of academic performance: A multinomial logit approach},
  author={Park, Kang H and Kerr, Peter M},
  journal={The Journal of Economic Education},
  volume={21},
  number={2},
  pages={101--111},
  year={1990},
  publisher={Taylor \& Francis}
}

@article{talib2012determinants,
  title={Determinants of Academic Performance of University Students.},
  author={Talib, Nadeem and Sansgiry, Sujit S},
  journal={Pakistan Journal of Psychological Research},
  volume={27},
  number={2},
  year={2012},
  publisher={Citeseer}
}

@article{wai2024most,
  title={The most successful and influential Americans come from a surprisingly narrow range of ‘elite’educational backgrounds},
  author={Wai, Jonathan and Anderson, Stephen M and Perina, Kaja and Worrell, Frank C and Chabris, Christopher F},
  journal={Humanities and Social Sciences Communications},
  volume={11},
  number={1},
  pages={1--10},
  year={2024},
  publisher={Palgrave}
}

@article{koech2016factors,
  title={Factors influencing career choices among undergraduate students in public universities in Kenya: A case study of university of Eldoret},
  author={Koech, Julius and Bitok, Jacob and Rutto, Daniel and Koech, Samson and Okoth, Joseph Onyango and Korir, Betty and Ngala, Hassan},
  journal={International Journal of Contemporary Applied Sciences},
  volume={3},
  number={2},
  pages={50--63},
  year={2016}
}

@article{mishkin2016career,
  title={Career choice of undergraduate engineering students},
  author={Mishkin, Hagit and Wangrowicz, Niva and Dori, Dov and Dori, Yehudit Judy},
  journal={Procedia-social and behavioral sciences},
  volume={228},
  pages={222--228},
  year={2016},
  publisher={Elsevier}
}

@article{quimby2006influence,
  title={The influence of role models on women's career choices},
  author={Quimby, Julie L and De Santis, Angela M},
  journal={The Career Development Quarterly},
  volume={54},
  number={4},
  pages={297--306},
  year={2006},
  publisher={Wiley Online Library}
}

@article{casanova2018factors,
  title={Factors that determine the persistence and dropout of university students},
  author={Casanova, Joana R and Cervero, Antonio and N{\'u}{\~n}ez, Jos{\'e} Carlos and Almeida, Leandro S and Bernardo, Ana},
  year={2018},
  publisher={Colegio Oficial De Psicologos Del Principado De Asturias}
}

@article{maluenda2022early,
  title={Early and dynamic socio-academic variables related to dropout intention: A predictive model made during the pandemic},
  author={Maluenda-Albornoz, Jorge and Infante-Villagr{\'a}n, Valeria and Galve-Gonz{\'a}lez, Celia and Flores-Oyarzo, Gabriela and Berr{\'\i}os-Riquelme, Jos{\'e}},
  journal={Sustainability},
  volume={14},
  number={2},
  pages={831},
  year={2022},
  publisher={Mdpi}
}

@article{gurak1992migration,
  title={Migration networks and the shaping of migration systems},
  author={Gurak, Douglas T and Caces, Fee},
  journal={International migration systems: A global approach},
  volume={150},
  pages={176},
  year={1992},
  publisher={Clarendon Press Oxford}
}

@article{santosa2021classification,
  title={Classification and prediction of students’ GPA using K-means clustering algorithm to assist student admission process},
  author={Santosa, Raden Gunawan and Lukito, Yuan and Chrismanto, Antonius Rachmat},
  journal={J. Inf. Syst. Eng. Bus. Intell},
  volume={7},
  number={1},
  pages={1},
  year={2021}
}

@misc{Desplaza60:online,
author = {Patricio Rodríguez},
title = {Desplazamiento de los estudiantes en Chile: implicancias para las políticas de salud y educación en el contexto del COVID-19 - CIPER Chile},
howpublished = {\url{https://www.ciperchile.cl/2020/04/28/desplazamiento-de-los-estudiantes-en-chile-implicancias-para-las-politicas-de-salud-y-educacion-en-el-contexto-del-covid-19/}},
month = {},
year = {2020},
note = {(Accessed on 09/14/2024)}
}

@article{oyelade2010application,
  title={Application of k Means Clustering algorithm for prediction of Students Academic Performance},
  author={Oyelade, Olanrewaju Jelili and Oladipupo, Olufunke O and Obagbuwa, Ibidun Christiana},
  journal={arXiv preprint arXiv:1002.2425},
  year={2010}
}

@article{odoardi2021can,
  title={Can social support compensate for missing family support? An examination of dropout rates in Italy},
  author={Odoardi, Iacopo and Furia, Donatella and Cascioli, Piera},
  journal={Regional Science Policy \& Practice},
  volume={13},
  number={1},
  pages={121--139},
  year={2021},
  publisher={Wiley Online Library}
}

@article{rodriguez2024does,
  title={Does private education pay off?},
  author={Rodr{\'\i}guez-Pose, Andr{\'e}s and Henry de Frahan, Rosalie},
  journal={The Annals of Regional Science},
  pages={1--26},
  year={2024},
  publisher={Springer}
}

@article{faggian2009universities,
  title={Universities, agglomerations and graduate human capital mobility},
  author={Faggian, Alessandra and McCann, Philip},
  journal={Tijdschrift voor economische en sociale geografie},
  volume={100},
  number={2},
  pages={210--223},
  year={2009},
  publisher={Wiley Online Library}
}

@article{faggian2018interregional,
  title={The interregional migration of human capital and its regional consequences: a review},
  author={Faggian, Alessandra and Rajbhandari, Isha and Dotzel, Kathryn R},
  journal={Transitions in Regional Economic Development},
  pages={227--256},
  year={2018},
  publisher={Routledge}
}

@article{ren2021family,
  title={Family socioeconomic status, educational expectations, and academic achievement among Chinese rural-to-urban migrant adolescents: The protective role of subjective socioeconomic status},
  author={Ren, Yi and Zhang, Feng and Jiang, Ying and Huang, Silin},
  journal={The Journal of Early Adolescence},
  volume={41},
  number={8},
  pages={1129--1150},
  year={2021},
  publisher={SAGE Publications Sage CA: Los Angeles, CA}
}

@article{rodriguez2020socio,
  title={Socio-economic status and academic performance in higher education: A systematic review},
  author={Rodr{\'\i}guez-Hern{\'a}ndez, Carlos Felipe and Cascallar, Eduardo and Kyndt, Eva},
  journal={Educational Research Review},
  volume={29},
  pages={100305},
  year={2020},
  publisher={Elsevier}
}

@article{diemer2020charting,
  title={Charting how wealth shapes educational pathways from childhood to early adulthood: a developmental process model},
  author={Diemer, Matthew A and Marchand, Aixa D and Mistry, Rashmita S},
  journal={Journal of Youth and Adolescence},
  volume={49},
  pages={1073--1091},
  year={2020},
  publisher={Springer}
}

@article{drager2022role,
  title={The role of parental wealth in children’s educational pathways in Germany},
  author={Dr{\"a}ger, Jascha},
  journal={European Sociological Review},
  volume={38},
  number={1},
  pages={18--36},
  year={2022},
  publisher={Oxford University Press}
}

@article{tuckman1970determinants,
  title={Determinants of college student migration},
  author={Tuckman, Howard P},
  journal={Southern Economic Journal},
  pages={184--189},
  year={1970},
  publisher={JSTOR}
}

@article{koch2023role,
  title={The role of immigrants, emigrants and locals in the historical formation of European knowledge agglomerations},
  author={Koch, Philipp and Stojkoski, Viktor and Hidalgo, C{\'e}sar A},
  journal={Regional Studies},
  pages={1--15},
  year={2023},
  publisher={Taylor \& Francis}
}

\end{document}